\documentclass[
twocolumn,
]{ceurart_custom}

\usepackage{listings}
\usepackage{pdfpages}

\begin{document}

\copyrightyear{2022}
\copyrightclause{Copyright for this paper by its authors.
  Use permitted under Creative Commons License Attribution 4.0
  International (CC BY 4.0).}

\conference{IJCAI 2023 Workshop on Deepfake Audio Detection and Analysis (DADA 2023), August 19, 2023, Macao, S.A.R}

\title{The defender's perspective on automatic speaker verification: An overview}



\address{\vspace{-3.5mm}}
\address[1]{Graduate Institute of Communication Engineering, National Taiwan University}
\address[2]{Department of Systems Engineering \& Engineering Management, The Chinese University of Hong Kong \vspace{-4mm}}

\author[1]{Haibin Wu}[
email=f07921092@ntu.edu.tw,
]

\author[2]{Jiawen Kang}[
]

\author[2]{Lingwei Meng}[
]

\author[2]{Helen Meng}[
]

\author[1]{Hung-yi Lee}[
]



\begin{abstract}
Automatic speaker verification (ASV) plays a critical role in security-sensitive environments. 
Regrettably, the reliability of ASV has been undermined by the emergence of spoofing attacks, such as replay and synthetic speech, as well as adversarial attacks and the relatively new partially fake speech. 
While there are several review papers that cover replay and synthetic speech, and adversarial attacks, there is a notable gap in a comprehensive review that addresses defense against adversarial attacks and the recently emerged partially fake speech. 
Thus, the aim of this paper is to provide a thorough and systematic overview of the defense methods used against these types of attacks.
\vspace{-1.5mm}
\end{abstract}


\maketitle


\textit{\vspace{-14.3mm}}

\section{Introduction}
\label{sec: introduction} 
\vspace{-2.5mm}
The past few years have witnessed significant advances in ASV, and this technique is now widely integrated into daily life, including voice activation in smartphones and e-banking authentication.
However, ASV is serious vulnerable to malicious spoofing attacks includes tactics such as replay and synthetic speech, adversarial attacks and recently emerged partially fake speech.

While there are several review papers that cover replay and synthetic speech \cite{yamagishi2021asvspoof,todisco2019asvspoof,kinnunen2017asvspoof,wu2015asvspoof}, and adversarial attacks \cite{tan2022adversarial}, there is a notable gap in a comprehensive review that addresses defense methods against adversarial attacks and the recently emerged partially fake speech.
The objective of this thesis is to provide a thorough and systematic overview of the defense methods used against these two types of attacks.
It is hoped that they will inspire further researches within the ASV community.

\textit{\vspace{-7mm}}
\section{Attacks \textit{\vspace{-2.5mm}}}
\label{section:intro spoofing attacks}

\subsection{Partially fake speech \textit{\vspace{-1.5mm}}}

\begin{figure}[ht]
  \centering
  \centerline{\includegraphics[width=1.0\linewidth]{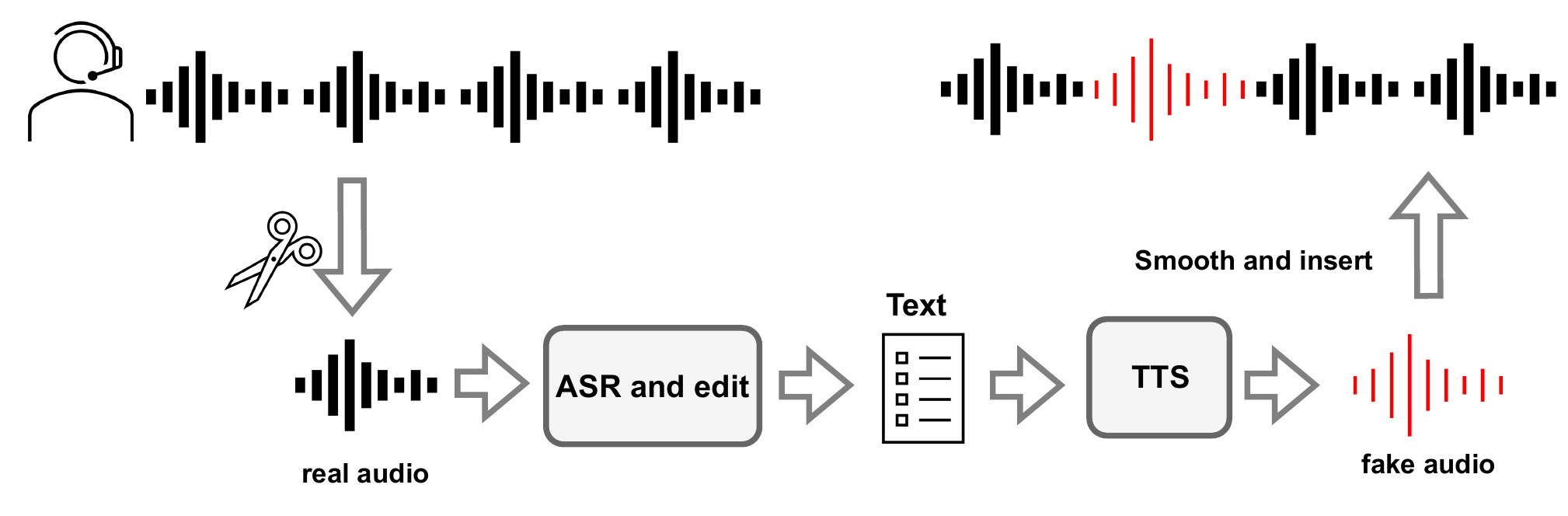}}
  \vspace{-0.2cm}
  \caption{The partially fake audio generation process. A small clip is selected from the user's utterance, the content is recognized using Automatic Speech Recognition (ASR), and the recognized content is modified to manipulate the meaning of the entire speech. The fake clip is then generated using Text-to-Speech (TTS) or Voice Conversion (VC), and inserted into the genuine utterance to generate the partially fake speech.}
  \label{fig:partially fake audio generation}
  \textit{\vspace{-7.5mm}}
\end{figure}

The first Audio Deep Synthesis Detection challenge (ADD 2022) \cite{Yi2022ADD} releases a kind of brand new attack, known as the partially fake speech attack \cite{yi2021half}.
The ASVspoof challenge \cite{yamagishi2021asvspoof,todisco2019asvspoof,kinnunen2017asvspoof,wu2015asvspoof} focuses on generating spoofing speech in its entirety, ignoring the scenario of partially fake speech, where small fake clips are hidden within a piece of real speech.
The generation of partially fake audio involves the insertion of only small clips of synthetic speech into the real speech as shown in Figure~\ref{fig:partially fake audio generation}, resulting in even more stealthy fake speech containing a significant amount of the genuine user's audio. 

Previous studies \cite{yi2021half,zhang2021initial} have shown that it is challenging to differentiate between partially fake and genuine audios by directly using existing state-of-the-art countermeasure models fostered by the ASVspoof challenge \cite{yamagishi2021asvspoof,todisco2019asvspoof,kinnunen2017asvspoof,wu2015asvspoof}.
These countermeasure models address the problem of identifying whether an entire audio utterance is genuine or fabricated. 
However, they are not equipped to identify anomalous regions within a single utterance.
\textit{\vspace{-3mm}}

\subsection{Adversarial attacks \textit{\vspace{-1.5mm}}}

\begin{figure}[ht]
  \centering
  \centerline{\includegraphics[width=0.7\linewidth]{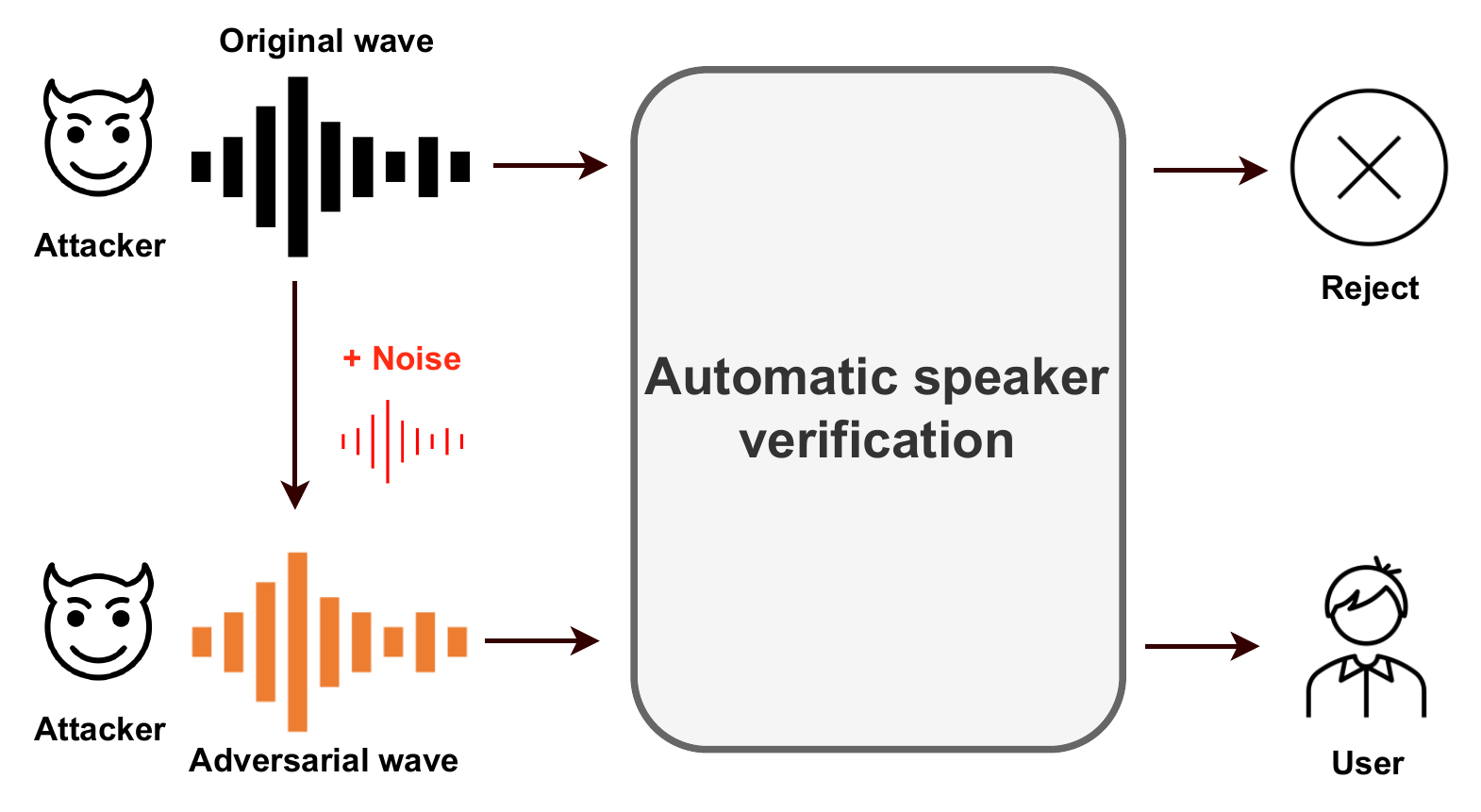}}
  \vspace{-0.2cm}
  \caption{A tiny adversarial noise is added to the original wave to get the adversarial one to fool the ASV falsely accept.}
  \label{fig:adv attack}
  \textit{\vspace{-5mm}}
\end{figure}

Speaker verification models are also subject to adversarial attacks \cite{abdullah2021sok,tan2022adversarial,das2020attacker} as shown in Figure~\ref{fig:adv attack}.
Kreuk et al. \cite{kreuk2018fooling} are among the pioneers in studying the susceptibility of ASV models to adversarial attacks.
Additionally, even the current state-of-the-art ASV models, including i-vector \cite{li2020adversarial} and x-vector \cite{villalba2020x} systems, are not immune to adversarial attacks.
\cite{liu2019adversarial} conducts a pioneering effort in exposing the adversarial weakness of countermeasure models and \cite{zhang2020black} further enhances the transferability of adversarial attacks through model ensemble. 

\textit{\vspace{-5mm}}
\section{Defense methods\textit{\vspace{-2.5mm}}}
\subsection{Tackle partially fake speech attacks \textit{\vspace{-3mm}}}
\begin{figure}[ht]
  \centering
  \centerline{\includegraphics[width=1.0\linewidth]{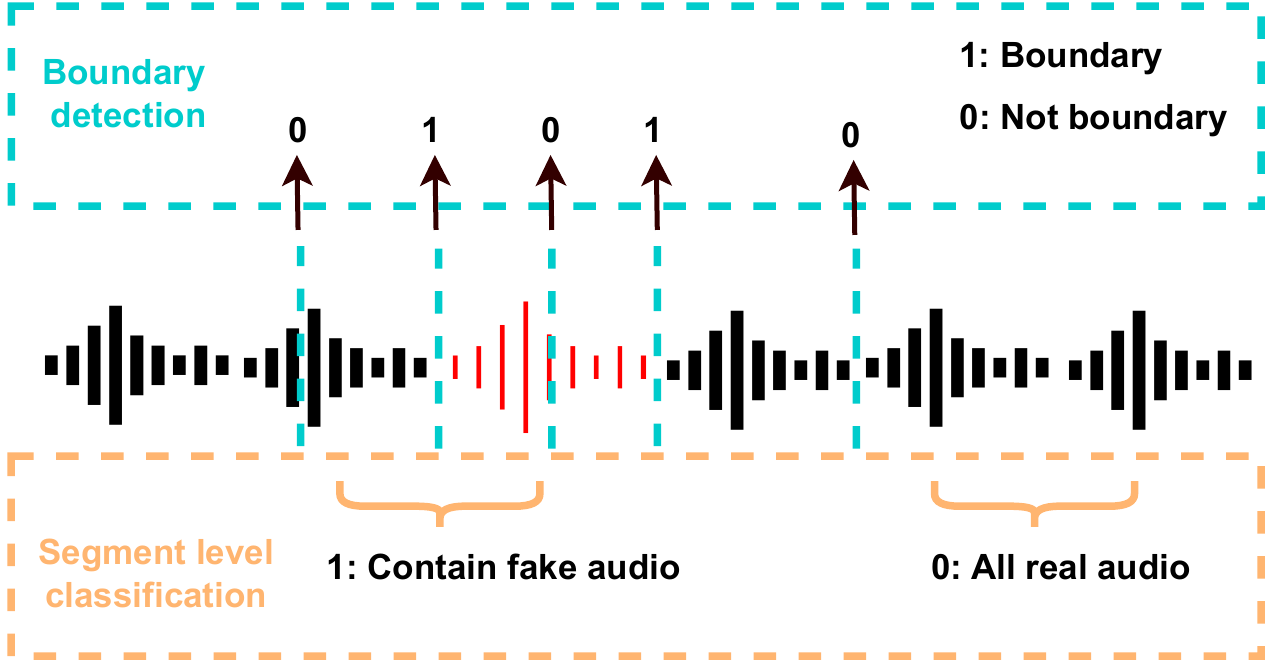}}
  \caption{The two categories of methods to tackle partially fake speech attacks. The black and red parts of the utterance are real and fake, respectively. The first approach, illustrated in the blue block, focuses on detecting the transition boundaries between the genuine and fake segments. The second method, depicted in the orange block, endeavors to distinguish between genuine and fake short segments.}
  \label{fig:two kinds of PF detection}
  \textit{\vspace{-5mm}}
\end{figure}
\noindent
Partially fake speech attacks are generated as shown in Figure~\ref{fig:partially fake audio generation}.
As this kind of attack is brand new, there have been only a few initiatives to handle this attack, and we categorize these efforts into two categories as shown in Figure~\ref{fig:two kinds of PF detection}: transition boundary detection \cite{wu2022partially,cai2022waveform,wang2022synthetic} and segment level classification \cite{yi2021half,zhang2021initial,zhang2021multi,zhang2022partialspoof}.
\textit{\vspace{-3mm}}

\subsubsection{SSL-based feature extractor\textit{\vspace{-1.5mm}}} 
Before delving into the two main approaches, let's first examine the feature engineering aspect of the task. 
Lv et al. \cite{lv2022fake} are the pioneers in utilizing self-supervised learning (SSL) models to tackle partially fake speech attacks.
Rather than using traditional acoustic features, they instead adopt XLS-R \cite{babu2021xls}, a self-supervised learning model, as the feature extractor.
Their method \cite{lv2022fake}, which involved simply adding a lightweight prediction head on top of the XLS-R model and fine-tuning the large XLS-R model, ultimately achieved first place out of 33 international teams in the ADD challenge \cite{Yi2022ADD}.

Their efforts \cite{lv2022fake} have taught us a valuable lesson - the acoustic features extracted by a fine-tuned self-supervised learning model can be incredibly helpful for detecting partially fake speech.
It's worth noting that the two main approaches introduced below can also harness the power of self-supervised learning models, provided there are sufficient computing resources available.
\textit{\vspace{-3mm}}

\subsubsection{Transition boundary detection \textit{\vspace{-1.5mm}}}
\cite{wu2022partially} is the first to introduce the transition boundary detection task for partially fake audio detection. 
The transition boundaries contain artifacts, such as discontinuity in speech and inconsistencies in ambient noise.
Inspired by the extraction-based question-answering models \cite{allam2012question} used in natural language processing (NLP), we refer to the boundary detection task as a question-answering or fake span discovery proxy task. 
In this task, the model is required to answer the question ``where is the fake clip?" in a piece of partially fake audio.
Extraction-based question-answering models in NLP typically take a question and a passage as input, construct representations for the passage and the question, match the question and passage embeddings, and then output the start and end positions of the answer within the passage.
In our case, the passage is the partially fake utterance, and the answer is the start and end time of the fake clip.
As depicted in the blue block of Figure~\ref{fig:two kinds of PF detection}, when the model is presented with a boundary frame between a real (black) and a fake (red) clip, it should predict ``1".
Conversely, when the model is presented with a non-boundary frame, it should predict ``0".
By training the model on the question-answering proxy task, the model can learn to find the concatenation boundaries with discontinuity and identify fake clips within an utterance, thus improving its ability to distinguish between audios with and without fake clips.
The proposed method placed the second out of 33 international teams in the ADD challenge \cite{Yi2022ADD}, even without the assistance of self-supervised learning features.

Wang et al. \cite{wang2022synthetic} divide the entire utterance into several chunks, and extracted acoustic features from each chunk to feed into the deep learning model.
The model is then tasked with determining whether a boundary exists within the given chunk by predicting ``1" if the chunk contains a boundary, or a ``0" if it does not.
Through training, the model gains the ability to identify clues such as speech discontinuity or inconsistencies in ambient noise, allowing it to effectively highlight potential boundaries.

Cai et al. \cite{cai2022waveform} propose to introduce the self-supervised learning model for frame-level boundary detection to detect partially fake speech.
They modify the method in \cite{wu2022partially} to further boost the detection performance:
1). Instead of solely focusing on transition boundaries that indicate inconsistency and discontinuity, \cite{cai2022waveform} proposes setting nearby frames of the boundaries as boundaries to increase robustness.
2). \cite{cai2022waveform} employs wav2vec 2.0 \cite{baevski2020wav2vec}, a self-supervised learning model as feature extractor and also fine-tunes the feature extractor during training. Utilizing the features from wav2vec 2.0 improves the performance by a relative 58.25\% compared to traditional acoustic features extracted by digital signal processing front-ends.

The main takeaway message from this subsection is that the transition boundaries can serve as a useful cue to identify partially fake audio, as it indicates discontinuity and inconsistency in speech. 
By tasking models with detecting these boundaries, they can learn to identify these cues and detect partially fake speech.
\textit{\vspace{-3mm}}

\subsubsection{Segment level classification \textit{\vspace{-1.5mm}}}
The goal of segment level classification is to distinguish between genuine and fake segments.
The short segments have different time resolutions, ranging from 1 frame (around 20 ms) to the entire utterance.
Segments that only contain genuine speech will be labeled as ``1", while all other segments will be labeled as ``0" as shown in the orange block of Figure~\ref{fig:two kinds of PF detection}.
Zhang et al. \cite{zhang2021initial} do the initial attempt to conduct segment level classification for partially fake speech detection with a fixed time resolution. 
In their subsequent works \cite{zhang2021multi}, they propose to train the countermeasure model by both the utterance level classification and segment level classification.
To further boost the countermeasure's performance, they \cite{zhang2022partialspoof} introduce the self-supervised learning models \cite{chen2022wavlm,baevski2020wav2vec} as the front-end feature extractor, and enable the model to learn segment level classification with different time resolutions, ranging from 1 frame to the entire utterance.

The time resolution used in segment level classification is a crucial hyperparameter for training. 
If the segment's frame number is too small, the model may not extract enough information to distinguish between genuine and fake segments. 
On the other hand, if the frame number is too large, the proportion of fake frames may be too small, resulting in fake frames being dominated by genuine frames.
Enabling the model learn from different time resolutions \cite{zhang2022partialspoof} is a reasonable solution to bypass the hyperparameter search.
Note that in Figure~\ref{fig:partially fake audio generation}, the inserted red clip can be from other genuine users. 
The segment level classification \cite{zhang2021initial,zhang2021multi,zhang2022partialspoof} does not consider this condition into account as in their produced dataset, the inserted clips are always fake.
\textit{\vspace{-3mm}}

\begin{figure*}
    \centering
    \includegraphics[width=16cm]{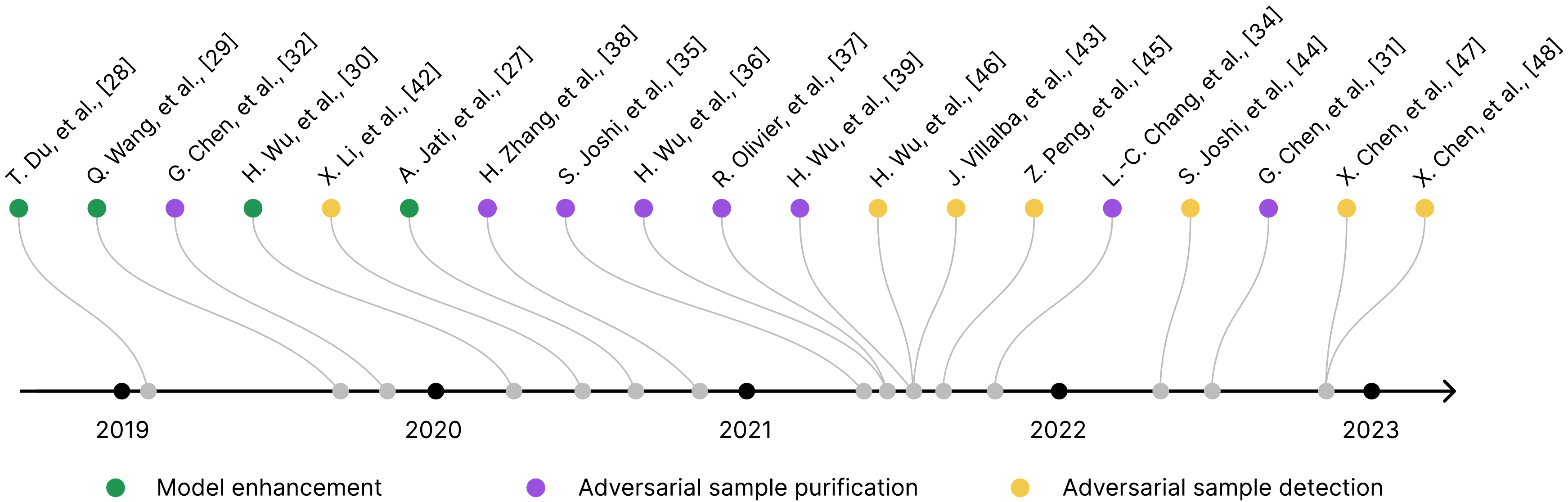}
    \vspace{-1cm}
    \caption{The timeline of defense methods for ASV against adversarial attacks. \textit{\vspace{-1.2cm}}}
    \label{fig:timeline_defense_adv}
\end{figure*}

\subsection{Defense against adversarial attacks \textit{\vspace{-1.5mm}}}
We propose to classify the defense methods into three categories and the timeline for related works is shown in Figure~\ref{fig:timeline_defense_adv}.
1). Model enhancement focuses on developing robust models during the training phase by modifying the models' internals, making the attackers difficult to find effective adversarial examples.
2). Adversarial sample purification aims to alleviate the superficial adversarial noise and transform adversarial samples into genuine samples.
3). Adversarial sample detection aims to distinguish between adversarial and genuine samples, allowing the identification and removal of adversarial samples.
\textit{\vspace{-3mm}}

\subsubsection{Model enhancement \textit{\vspace{-1.5mm}}}
\cite{jati2021adversarial,du2020sirenattack,wang2019adversarial} adopt adversarial training to alleviate the vulnerability of ASV against adversarial attacks.
Wu et al. \cite{wu2020defense} also investigate improving the adversarial robustness for countermeasures by adversarial training.

Model enhancement methods involve modifying the model's parameters, and they can usually work together with purification and detection methods.
\textit{\vspace{-3mm}}

\subsubsection{Adversarial sample purification \textit{\vspace{-1.5mm}}}
Previous efforts for purification can be classified into 5 categories: Lossy pre-processing, adding noise, generative method, denoising method and filtering.

The ``Lossy pre-processing" approach treats adversarial perturbations as redundant information and discards it to improve the model's adversarial robustness. 
Chen et al. \cite{chen2022towards} consider adversarial perturbations as redundant information and use lossy speech compression techniques to mitigate these perturbations.
Quantization \cite{chen2019real,chen2022towards} involves rounding each audio sample point to the nearest integer multiple of a factor $q$, which can impact the fragile adversarial perturbations.
Chen et al. \cite{chen2022towards} propose to do k-means \cite{hartigan1979algorithm} on the acoustic features to get clusters of acoustic features, and use the clusters to represent the acoustic features.

The ``adding noise" approach aims to disrupt and neutralize adversarial perturbations, by introducing additional noise, typically Gaussian.
Randomized smoothing \cite{chen2019real,chang2021defending,chen2022towards,joshi2021adversarial} involves adding random Gaussian noise to the input utterances before sending them to the ASV to counter the adversarial perturbations. 
\cite{wu2021voting} adopts to the idea of ``voting for the right answer" to prevent risky decisions of ASV in blind spot areas. To achieve this, they samples the neighbors of a given utterance by random sampling using Gaussian noise, and allow the neighbors to vote on whether the utterance should be accepted by the ASV model or not, rather than relying solely on the prediction of the single utterance. Olivier et al. \cite{olivier2021high} is an enhanced version by adding Gaussian noise to the high-frequency region rather than the entire utterance.

The ``Denosing method" treats adversarial noise as a specific kind of noise and aims to estimate and eliminate it.
Chang et al. \cite{chang2021defending} suggest using a denoising algorithm tailored for Gaussian noise and they contend that the denoising algorithm can also cleanse the adversarial noise.
Zhang et al. \cite{zhang2020adversarial} propose to employ an adversarial separation network, which is trained using the adversarial-genuine data pairs, to estimate and purify the adversarial noise. This method requires prior knowledge of adversarial sample generation.

The ``generative method" approach typically involve training a generative model to model the genuine data manifolds and using this model to pull the adversarial samples towards the genuine data manifolds.
Wu et al. \cite{wu2021improving} propose the SSLM-based reconstruction to alleviate the superficial adversarial noise and maintain key information for genuine samples. They \cite{wu2021improving} utilize the self-supervised learning models to extract key features from the adversarial samples, and do reconstruction to pull the inputs to the genuine data manifold.
Joshi et al. \cite{joshi2021adversarial} use the encoder of a VAE \cite{kingma2013auto} to project testing data onto a latent posterior that aligns with the genuine manifold. They then use the decoder to re-generate the input data based on the hidden embedding sampled by the latent posterior, thereby purifying superficial adversarial noise.
Joshi et al. \cite{joshi2021adversarial} borrow the DefenseGAN from computer vision \cite{samangouei2018defense}. The DefenseGAN projects the testing data, either adversarial or genuine, into the low-dimensional manifold of genuine data to get the hidden embeddings and then re-generate the testing data by the generator using such embeddings.

``Filtering", also known as local smoothing, helps smooth and alleviate the superficial adversarial perturbations. Local smoothing involves applying Gaussian, mean, and median filters to the waveform to purify the adversarial noise. \cite{chen2019real,chen2022towards} and \cite{wu2020defense} utilize local smoothing to defend ASV and countermeasures, respectively.
\textit{\vspace{-3mm}}

\subsubsection{Adversarial sample detection \textit{\vspace{-1.5mm}}}

The detection methods can be classified into two categories based on whether they require prior knowledge about adversarial sample generation: attack-dependent or attack-independent detection methods.

The attack-dependent methods usually leverage the deep learning models to implicitly find cues to differentiate between specific kinds of adversarial samples and genuine samples using both adversarial and genuine data. 
Li et al. \cite{li2020investigating} propose to train a detector using the binary classification loss to distinguish the adversarial and genuine samples. They find their detector is unable to detect unseen adversarial samples derived by other adversarial attack algorithms that are not used during training.
Based on that different kinds of adversarial samples attain different attack signatures, Villalba et al. \cite{villalba2021representation} propose to train an x-vector \cite{villalba2020x} system to extract the bottleneck features as the attack signatures using various types of adversarial samples. After training the x-vector system, attack signatures will be extracted for different types of attacks. During inference, the testing utterance is inputted, and the x-vector feature extractor will extract the hidden embeddings. These embeddings are then compared with the enrolled attack signatures to determine whether the testing utterance is an adversarial sample or not. To further improve the performance of the attack signature extractor, Joshi et al. \cite{joshi2022advest} propose training the attack signature extractor using adversarial perturbations instead of adversarial examples. They argue that the adversarial perturbations eliminate redundant information from the adversarial samples. They then train an adversarial perturbation estimator to extract adversarial perturbations from the input utterance and use the attack signature extractor to extract hidden features to detect the adversarial samples.

Attack-independent methods treat the detection of adversarial samples as an anomaly detection problem. Genuine data samples always exhibit some properties that are absent or different for adversarial samples. Therefore, attack-independent detection methods can exploit the inconsistency of these internal properties to distinguish between adversarial and genuine samples.
Wu et al. \cite{wu2021improving} leverage the ASV score difference before and after putting the testing utterance into SSLMs as an indicator to differentiate between adversarial and genuine samples. Specifically, for genuine samples, the ASV score difference before and after putting the utterance into SSLMs is small, while for adversarial samples, the difference is large.
Peng et al. \cite{peng2021pairing} propose to detect adversarial samples using twin ASV models, including one premier model that is exposed to attackers and is fragile under adversarial attack, and one mirror model that is robust to adversarial attacks and cannot be accessed by attackers. When a genuine sample is inputted, both the premier and mirror models produce similar predictions. However, when an adversarial sample is inputted, the models produce different predictions. Peng et al. \cite{peng2021pairing} leverage the score inconsistency between genuine and adversarial samples to detect adversarial samples.
Wu et al. \cite{wu2022adversarial} utilize the vocoders to re-synthesize the input utterance and find that the difference between the ASV scores for the original and re-synthesized utterance is a good indicator for discrimination between genuine and adversarial samples. To be specific, the score difference for adversarial samples is large, while it is small for genuine samples.
Chen et al. \cite{chen2022masking} utilize two kinds of hand-crafted masks to detect adversarial samples: they mask parts of the input speech features. They claim the masked parts contain less speaker information and won't affect the ASV scores for genuine samples two much, but will greatly impact the adversarial samples. By comparing the absolute difference of scores before and after masking, they are able to detect adversarial examples. The two masks used are MLFB-H, which masks the high frequencies of LogFBank, and MLFB-D, which masks the time-frequency bins whose absolute values of their one-order difference along the frequency axis are smaller than a threshold. Chen et al. \cite{chen2022lmd} further enhance the detection performance by learning such mask matrix by a deep recurrent networks, rather than using hand-crafted masks.

\textit{\vspace{-7mm}}
\section{Future directions \vspace{-2.5mm}}
\label{sec:future directions}
For the future directions of partially fake speech attacks:
1). Data collection. The collection of data is a crucial component in developing an effective defense system against partially fabricated speech. Only 100k utterances are collected by \cite{Yi2022ADD} for partially fake detection and the transition boundaries are not stealthy enough. To this end, there exists a pressing need to investigate the generation of more data with discreet transition boundaries, while carefully considering the linguistic and acoustic characteristics involved. This undertaking is of great significance and warrants further exploration.
2). Reduce training efforts. The state-of-the-art (SOTA) methodology for partially fake speech detection involves the fine-tuning of the entire SSLMs. The SSLM in \cite{lv2022fake} is with 2 billion parameters, which presents a challenge for academic researchers when attempting to fine-tune the model. Several works have emerged that offer promising avenues for minimizing training efforts while maximizing the benefits of SSLMs, including linear probing, adapter, and prompt techniques. Exploring these approaches may significantly enhance the efficiency of adopting SSLMs for partially fake speech detection.
3). Model compression. The current state-of-the-art detection method relies heavily on large-scale SSLMs. The parameter number of the SSLM used in \cite{lv2022fake} is 2 billion parameters. Therefore, investigating approaches to reduce the model size is a crucial research endeavor. This issue warrants considerable attention as it has significant implications for the scalability, computational efficiency, and generalizability of partially fake speech detection systems.

The re-synthesis-based adversarial sample detection methods achieves the SOTA \cite{wu2022adversarial,chen2022masking,chen2022lmd}.
An effective audio re-synthesis method for adversarial sample detection must possess two critical properties. Firstly, the score variations between the original and re-synthesized utterances should be minimal for genuine samples. Secondly, the score variations between the original and re-synthesized utterances for adversarial samples should be substantial. Investigating approaches for refining the design of audio re-synthesis methods to further optimize these properties represents a valuable research direction. By enhancing the efficacy of the audio re-synthesis method, it would be possible to improve the reliability and accuracy of detection systems.

\vspace{-4mm}

\section{Conclusion \vspace{-2.5mm}}

This paper reviews the defense methods against adversarial attacks and partially fake speech attacks that have recently emerged.
We hope the comprehensive review and comparisons can inspire future works to boost the robustness of ASV.
Further investigation is needed to explore future directions as in Section~\ref{sec:future directions}

\renewcommand\refname{References \vspace{-2mm}}
\bibliography{reference}

\end{document}